\documentclass[aps,prd,showpacs,showkeys,amsfonts,amsmath,reprint,floatfix]{revtex4-1}  %final

\usepackage[utf8]{inputenc}
\usepackage[T1]{fontenc}
\usepackage{hyperref}
\usepackage{bm}
\usepackage{graphicx}
\usepackage{booktabs}
\usepackage[separate-uncertainty=true]{siunitx}
\usepackage{amssymb}
\usepackage[figuresright]{rotating}
\usepackage{mathtools}

\newcommand{\dd}{\mathrm{d}}
\newcommand{\abs}[1]{\lvert#1\rvert}

\newcolumntype{d}[1]{D{.}{.}{#1} }
\DeclareMathOperator{\Real}{Re}
\DeclareMathOperator{\Imaginary}{Im}

\begin{document}

\title{Search for anisotropic Lorentz invariance violation with $\gamma$-rays}
\date{\today}

\author{Fabian Kislat}
\email[Send correspondence to: ]{fkislat@physics.wustl.edu}
\author{Henric Krawczynski}
\affiliation{Washington University in St.\ Louis, Department of Physics and McDonnell Center for the Space Sciences, St.\ Louis, MO 63105}

\pacs{11.30.Cp, 95.85.Pw, 98.54.Cm}

\keywords{Lorentz invariance; Standard-Model Extension; AGN; Gamma-rays; Fermi}

\begin{abstract}
  While Lorentz invariance, the fundamental symmetry of Einstein's theory of General Relativity, has been tested to a great level of detail, Grand Unified Theories that combine gravity with the other three fundamental forces may result in a violation of Lorentz symmetry at the Planck scale.
  These energies are unattainable experimentally.
  However, minute deviations from Lorentz invariance may still be present at much lower energies.
  These deviations can accumulate over large distances, making astrophysical measurements the most sensitive tests of Lorentz symmetry.
  One effect of Lorentz invariance violation is an energy dependent photon dispersion of the vacuum resulting in differences of the light travel time from distant objects.
  The Standard-Model Extension (SME) is an effective theory to describe the low-energy behavior of a more fundamental Grand Unified Theory, including Lorentz and CPT violating terms.
  In the SME the Lorentz violating operators can in part be classified by their mass-dimension $d$, with the lowest order being $d=5$.
  However, measurements of photon polarization have constrained operators with $d=5$ setting lower limits on the energy at which they become dominant well beyond the Planck scale.
  On the other hand, these operators also violate CPT, and thus $d=6$ could be the leading order.
  In this paper we present constraints on all 25 real coefficients describing anisotropic non-birefringent Lorentz invariance violation at mass dimension $d=6$ in the SME.
  We used Fermi-LAT observations of 25 active galactic nuclei to constrain photon dispersion and combined our results with previously published limits in order to simultaneously constrain all 25 coefficients.
  This represents the first set of constraints on these coefficients of a mass-dimension $d=6$, whereas previous measurements were only able to constrain linear combinations of all 25 coefficients.
\end{abstract}

\maketitle

\section{Introduction}
Lorentz invariance is the fundamental symmetry of Einstein's theory of relativity.
It has been established by early experiments such as the Michelson-Morley experiment~\cite{aa_michelson_ew_morley_1887} and has since been verified to great precision~\cite{[][{ (2015).}]va_kostelecky_n_russell_2014}. % Cheating here to get the ``(2015)'' after the arXiv reference in the bibliography
However, unified theories of General Relativity and the Standard Model of particle physics suggest that Lorentz symmetry may be broken at the Planck energy scale ($E_P = \sqrt{c^5\hbar/G} \approx \SI{1.22e19}{GeV}$)~\cite{g_amelino_camelia_etal_1998,*lj_garay_1998,*r_gambini_j_pullin_1999,*d_mattingly_2005,*t_jacobson_etal_2006,*s_liberati_l_maccione_2009}.
Lorentz invariance violation has to be suppressed at lower energies, but tiny deviations may still exist, motivating sensitive tests of Lorentz invariance.

In the photon sector violations of Lorentz symmetry include vacuum dispersion and vacuum birefringence~\cite{va_kostelecky_m_mewes_2008}.
Even though these effects are suppressed at observable energies, $E \ll E_P$, astrophysical observations can still be sensitive to new physics since tiny deviations accumulate over large distances~\cite{g_amelino_camelia_etal_1998,*lj_garay_1998,*r_gambini_j_pullin_1999,*d_mattingly_2005,*t_jacobson_etal_2006,*s_liberati_l_maccione_2009}.
Vacuum dispersion can be tested using astrophysical time-of-flight measurements: when observing a time-variable or transient source at large redshift, tiny variations in the photon velocity will accumulate leading to differences in the arrival time of photons at different wavelengths.
Using Fermi LAT observations of Gamma-Ray Bursts, linear photon dispersion has been ruled out beyond the Planck scale~\cite{aa_abdo_etal_2009}.
Similarly, vacuum birefringence can be probed with astrophysical polarization measurements.
In this case, the effects of tiny deviations from an isotropic vacuum will accumulate over extragalactic distances resulting in a measurable rotation of the polarization plane of linearly polarized light as a function of energy. 
Typically, the strongest constraints on Lorentz invariance violation result from astrophyical polarization measurements.
This can be understood by the fact that in a dispersion study over the baseline $L$ the sensitivity is given by arrival time variations $\delta t \propto \delta v\,L$,  whereas in a polarimetric study, the sensitivity is determined by the phase difference $\delta\phi \propto \omega\,\delta v\,L$ with $\omega$ being the frequency of the light, resulting in an improvement in sensitivity of $1/\omega$ compared to time-of-flight measurements, see e.\,g.~\cite{va_kostelecky_m_mewes_2009}.
Owing the high sensitivity of polarization observations, constraints from time-of-flight measurements are most interesting for testing theories or constraining parameters which do not predict any vacuum birefringence.

The Standard-Model Extension (SME,~\cite{d_colladay_va_kostelecky_1997,*d_colladay_va_kostelecky_1998,*va_kostelecky_2004,va_kostelecky_m_mewes_2009}) is an effective field theory to describe the low-energy phenomenology of a high-energy theory and includes effects of General Relativity and the Standard Model of particle physics.
Furthermore, it allows one to introduce Lorentz invariance and CPT symmetry violating terms in the Lagrange density.
Interpreting constraints on Lorentz invariance violation in terms of limits on the coefficients of the SME has the advantage over model independent tests, that results from different kinds of experiments (e.\,g.~polarization and time-of-flight measurements) can be compared directly.
The disadvantage of this approach is that some models of quantum gravity, such as theories of Doubly-Special Relativity (DSR,~\cite{[][{ and references therein.}]g_amelino_camelia_2010}), cannot be described in the effective field theory framework (however, see Sec. IV.F.3 of Ref.~\cite{va_kostelecky_m_mewes_2009} for a critique of DSR).
The additional Lorentz and/or CPT violating terms in the action of the SME can be ordered by the mass-dimension of the corresponding operator.
Operators of dimension $d$ lead to a dispersion proportional to $(E/E_\text{Planck})^{d-4}$, meaning that the renormalizable operators of $d \leq 4$ are unsuppressed with respect to conventional physics.
Thus, it is obvious, that only the non-renormalizable Lorentz-violating terms of $d \geq 5$ may contribute.

One of the problems of the Standard Model is that radiative corrections due to particle interactions can create unsuppressed $d \leq 4$ Lorentz violating terms~\cite{j_collins_etal_2004}.
One of the attractive features of Supersymmetric (SUSY) theories is that spontaneous symmetry breaking can suppress these terms~\cite{j_collins_etal_2004,pa_bolokhov_etal_2005}.

As mentioned above, the leading order Lorentz-violating operators of dimension $d=5$ lead to a photon dispersion linear in energy, which has been constrained beyond the Planck scale in the isotropic case~\cite{aa_abdo_etal_2009}.
Furthermore, in the SME all operators of this mass-dimension also result in birefringence, and can, therefore, be constrained much more strongly by polarization measurements~\cite{va_kostelecky_m_mewes_2013,*k_toma_etal_2012,*p_laurent_etal_2011,*fw_stecker_2011}.
The next-to-leading order operators of $d=6$ are generally suppressed compared to the leading order operators.
However, $d=5$ operators not only violate Lorentz invariance but also break CPT symmetry.
Additionally, the above mentioned suppression of induced lower-dimension Lorentz-violating operators through SUSY breaking is not sufficient, and fine-tuning will be required.
On the other hand, the terms of mass-dimension $d=6$ conserve CPT, and induced dimension-4 terms are sufficiently suppressed in SUSY theories~\cite{j_collins_etal_2004,pa_bolokhov_etal_2005}.
Therefore, it may well be possible that the lowest order non-vanishing Lorentz invariance violating terms are of mass-dimension $d=6$.

In the $d=6$ case, there is a subset of $(d-1)^2 = 25$ non-birefringent Lorentz-violating operators, which cannot be constrained through polarization measurements.
This motivates a dedicated search for photon dispersion proportional to $E^2$.
In general, Lorentz invariance violation can lead to an anisotropic photon dispersion.
A spherical decomposition results in 25 real coefficients, which can be constrained by observing photon dispersion from at least 25 astrophysical sources distributed evenly on the sky.
So far, constraints on quadratic photon dispersion have been derived from the observation of 4 gamma-ray bursts (GRBs) by Fermi LAT~\cite{v_vasileiou_etal_2013}, one GRB observed by RHESSI~\cite{se_boggs_etal_2004}, and four flares of Active Galactic Nuclei (AGN) observed by the TeV gamma-ray telescopes H.E.S.S.~\cite{f_aharonian_etal_2008,*a_abramowski_etal_2011,a_abramowski_etal_2015}, MAGIC~\cite{j_albert_etal_2008}, and Whipple~\cite{sd_biller_etal_1999}.
Constraints on linear dispersion from SWIFT, HETE, and BATSE observations of GRBs~\cite{j_ellis_etal_2006,*j_ellis_etal_2008} could in principle be converted to limits on quadratic dispersion. 
However, due to the much lower energies probed in these cases, the resulting constraints are not competitive.

While the work presented here as well as in the above mentioned references, consider systematic effects of Lorentz invariance violation, it is expected that the foamy structure of spacetime in models of Quantum Gravity may lead to a stochastic variation of the velocity of photons of the same energy~\cite{wa_christiansen_etal_2006,*g_amelino_camelia_l_smolin_2009}.
In general, both 'stochastic' and 'systematic' Lorentz invariance violations may be present.
Recently, stochastic variations of the linear photon dispersion have been constrained at the Planck scale by Fermi observations of GRB090510~\cite{v_vasileiou_etal_2015}.

In this paper, we analyzed Fermi LAT data~\cite{wb_atwood_etal_2009} of 25 AGN, and derived limits on photon dispersion for all of them.
We combine these limits with the previously published results in order to derive limits on the complete set of non-birefringent Lorentz-violating coefficients with mass-dimension 6 in the Standard-Model Extension.

In section~\ref{sec:framework}, we summarize the theoretical foundation of our analysis in the SME.
In section~\ref{sec:methods}, we give a brief introduction to the Fermi LAT.
In the same section we also describe the \emph{DisCan} method, which we used to constrain photon dispersion from the individual AGN studied in this analysis.
Our source selection and data set is described in section~\ref{sec:sources}, and our results and the combination with previously published results will be presented in section~\ref{sec:limits}.
Finally, we summarize our findings in section~\ref{sec:summary}.

\section{Mathematical framework}\label{sec:framework}
The basic assumption of the Standard Model Extension is that the theoretical framework of the Standard Model and General Relativity is the low-energy limit of a unified quantum gravity theory, which holds at the Planck energy scale.
In an expansion approximating the full theory, the action of the Standard Model is the zeroth-order term.
The Standard Model Extension considers additional terms in the action, whose magnitude can be constrained by observational data.
These additional terms are ordered by the mass-dimension $d$ of the tensor operator, and operators with $d>4$ lead to Lorentz invariance violation.
Although the full theory is thought to be Lorentz invariant and consistent with the cosmological principle, Lorentz invariance and isotropy of space breaking terms can arise dynamically.

A general Lagrange density of the photon sector can be written as~\cite{va_kostelecky_m_mewes_2009}:
\begin{equation}
  \begin{split}
    \mathcal{L} = &-\tfrac{1}{4}F_{\mu\nu}F^{\mu\nu} + \tfrac{1}{2}\epsilon^{\kappa\lambda\mu\nu}A_\lambda(\hat{k}_{AF})_\kappa F_{\mu\nu} \\
      &-\tfrac{1}{4}F_{\kappa\lambda}(\hat{k}_F)^{\kappa\lambda\mu\nu}F_{\mu\nu},
  \end{split}
\end{equation}
where the differential operators $\hat{k}_{AF}$ are CPT-odd and only contain coefficients of odd mass-dimension, whereas the operators $\hat{k}_F$ only contain coefficients of even $d$ and are CPT-even.
The equation of motion is derived by varying the Lagrangian and the dispersion relation follows from the equation of motion.
In the vacuum case it can be written as:
\begin{equation}
  E(p) \simeq \bigl(1 - \varsigma^0 \pm \sqrt{(\varsigma^1)^2 + (\varsigma^2)^2 + (\varsigma^3)^2}\bigr) \, p.
\end{equation}
An expansion in mass-dimension and spherical decomposition yield for photons of momentum $p$ arriving from direction $(\theta_k, \varphi_k)$:
\begin{align}\label{eq:sigma0}
  \varsigma^0 &= \sum_{djm}p^{d-4} Y_{jm}(\theta_k, \varphi_k)c_{(I)jm}^{(d)}, \\
  \notag\varsigma^\pm &= \varsigma^1 \pm \varsigma^2 \\
      &= \sum_{djm}p^{d-4} \prescript{}{\mp2}{Y}_{jm}(\theta_k, \varphi_k) \left(k_{(E)jm}^{(d)} \mp ik_{(B)jm}^{(d)}\right), \\
  \varsigma^3 &= \sum_{djm}p^{d-4} Y_{jm}(\theta_k, \varphi_k)k_{(V)jm}^{(d)},
\end{align}
where $c_{(I)jm}^{(d)}$ represents sets of $(d-1)^2$ non-birefringent CPT-even coefficients, i.\,e. they are non-zero only for even $d$.
Furthermore, $d \geq 4$, $j = 0 \ldots d-2$, and $\abs{m} \leq j$.
While, $\varsigma^0$ only contains non-birefringent CPT-even coefficients, $\varsigma^3$ and the combinations $\varsigma^\pm = \varsigma^1 \mp i\varsigma^2$ contain birefringent CPT-even and birefringent CPT-odd coefficients, respectively.
Since Lorentz symmetry is well-established, any Lorentz violating effect has to be small.
It is therefore expected that coefficients of Lorentz invariance violating operators are suppressed by a large scale, typically a factor of $M_\text{Planck}^{d-4}$~\cite{va_kostelecky_m_mewes_2009}.
The coordinates $(\theta_k, \varphi_k)$ are in a Sun-centered celestial equatorial frame, such that $\theta_k = \SI{90}{\degree}-\delta_k$ and $\varphi_k=\alpha_k$, where $\alpha_k$ and $\delta_k$ are the right ascension and declination of the $k$th astrophysical source, respectively.

In the following, we will assume a CPT-even non-birefringent vacuum model, which has nonzero coefficients only in even dimensions $d$, with the leading order being $d=6$ (at dimension 4 the photon dispersion only depends on direction, not energy, and therefore cannot be measured with astrophysical observations).
Using the approximation $E \simeq p$ in Eq.~\eqref{eq:sigma0}, the operators of this mass-dimension lead to a photon dispersion that is quadratic in energy.
Thus, the difference of arrival times of two photons with energies $E_1$ and $E_2$ emitted simultaneously from an astrophysical source at redshift $z_k$ is given by
\begin{multline}\label{eq:dispersion}
  t_2 - t_1 \approx \int\limits_0^{z_k} \frac{v_1 - v_2}{H_z} \dd z \\
    \approx (E_2^2 - E_1^2) \int\limits_0^{z_k}\frac{(1 + z)^2}{H_z} \dd z \sum_{jm} Y_{jm}(\theta_k, \varphi_k) c_{(I)jm}^{(6)},
\end{multline}
where
\begin{equation}
  H_z = H_0[\Omega_r(1+z)^4 + \Omega_m(1+z)^3 + \Omega_k(1+z)^2 + \Omega_\Lambda]^{\frac{1}{2}}
\end{equation}
is the Hubble expansion rate at redshift $z$ with the present day Hubble constant $H_0 \simeq \SI{70}{km.s^{-1}.Mpc^{-1}}$, the radiation density $\Omega_r \simeq 0.015$, matter density $\Omega_m \simeq 0.27$, vacuum density $\Omega_\Lambda \simeq 0.73$, and curvature density $\Omega_k = 1-\Omega_r-\Omega_m-\Omega_\Lambda$~\cite{n_jarosik_etal_2011}.

Introducing the dispersion coefficient
\begin{equation}\label{eq:dispersion-theta}
  \vartheta_k = \int\limits_0^{z_k}\frac{(1 + z)^2}{H_z} \dd z \sum_{jm} Y_{jm}(\theta_k, \varphi_k) c_{(I)jm}^{(6)},
\end{equation}
equation~\eqref{eq:dispersion} can be written as
\begin{equation}
  \Delta t_k = \vartheta_k(E_2^2 - E_1^2)_k,
\end{equation}
where the index $k$ indicates the $k$th astrophysical source being studied.
With the redshift and light-travel-time weighted dispersion coefficient
\begin{equation}\label{eq:gamma_k}
  \gamma_k = \frac{\vartheta_k}{\int_0^{z_k}\frac{(1 + z)^2}{H_z} \dd z}
\end{equation}
one finds a system of equations to calculate the coefficients $c_{(I)jm}^{(6)}$:
\begin{equation}\label{eq:cI_system}
  \sum_{\substack{j = 0 \ldots 4 \\
                  m = -j \ldots j}}
    Y_{jm}(\theta_k, \varphi_k)c_{(I)jm}^{(6)} = \gamma_k.
\end{equation}
At leading order, $d=6$, there are 25 complex coefficients $c_{(I)jm}^{(6)}$.
However, since the $\gamma_k$ are real, the structure of the spherical harmonics leads to the reality condition
\begin{equation}\label{eq:reality_condition}
  c_{(I)j-m}^{(6)} = (-1)^m \bigl(c_{(I)jm}^{(6)}\bigr)^*,
\end{equation}
resulting in a total of 25 real coefficients, with all $c_{(I)j0}^{(6)}$ real.

Thus, measurements of $\gamma_k$ from at least 25 sources are required to constrain all coefficients individually.
If no significant deviation of photon travel times is found, positive and negative limits on $\gamma_k$ are determined constraining a volume in the 25-dimensional parameter space of the $c_{(I)jm}^{(6)}$.

\section{Instrument and Methods}\label{sec:methods}
\subsection{The Fermi LAT}
The Large Area Telescope (LAT) is the primary instrument on the Fermi Gamma-ray Space Telescope~\cite{wb_atwood_etal_2009}, covering the $\gamma$-ray energy band from~\SI{20}{MeV} to more than~\SI{300}{GeV}.
It is an imaging telescope with a wide field of view of~\SI{2.4}{sr} that covers the entire sky every two orbits, and as such is ideally suited for the study presented here because it allows us to obtain densely sampled long-term light curves of AGN.
The LAT is a pair-conversion telescope consisting of a converter-tracker and a calorimeter.
Gamma-rays convert in the tungsten layers of the converter-tracker, and the tracks of the $e^+e^-$ pair are recorded in silicon strip detectors in order to reconstruct the direction of the incident gamma-ray.
The electromagnetic shower initiated by the $e^+e^-$ pair is then absorbed in the calorimeter to measure the energy deposition and, thus, reconstruct the energy of the gamma-ray.
Above an energy of \SI{1}{GeV} the angular resolution of the LAT is better than~$1^\circ$.

\subsection{The \emph{D\lowercase{is}C\lowercase{an}} Method}\label{sec:discan}
As mentioned in the introduction, equations~\eqref{eq:dispersion} and \eqref{eq:dispersion-theta}, the Lorentz violating operators of mass-dimension $d=6$ lead to a quadratic dependence of the photon travel time from the source to the observer on the photon energy.
In case of a time-variable source, this will generally smear out the structure of the light curve.
In the dispersion cancellation method, this effect is corrected for, by adjusting the arrival time of each photon proportional to $-\vartheta E^2$:
\begin{equation}
  t'_0 = t_\mathrm{arr} - \vartheta(E^2 - \langle E^2\rangle),
\end{equation}
where $\langle E^2\rangle$ is the average $E^2$ of the observed photons.
One then finds the value of $\vartheta$ that leads to the ``least washed-out'' light curve.
In this way no binning in energy is necessary.

In the \emph{DisCan} method~\cite{jd_scargle_etal_2008} a binning of photons in time is furthermore avoided as follows.
Each photon at arrival time $t_i$ is assigned a time bin of width
\begin{equation}
  \Delta t_i = \frac{t_{i+1} - t_{i-1}}{2}.
\end{equation}
In order to represent the light curve in this way, the contents of each time bin are set to
\begin{equation}
  w_i = 1/\Delta t_i.
\end{equation}
This leads to an accurate albeit choppy representation of the light curve that does not require binning photons in time.
In order to reduce choppiness, we chose to combine each 10 consecutive photons into one wider time bin with appropriate weight.
Furthermore, we required that the last photon in each bin is separated from the next by more than $1\si{s}$, otherwise the bin is extended by one photon.
Thus the final duration of the $n$-th bin spanning photons $n_1 \ldots n_2$ is
\begin{equation}
  \Delta t_n = \frac{t_{n_2+1} - t_{n_2} - t_{n_1} + t_{n_1-1}}{2}
\end{equation}
with the weight
\begin{equation}
  w_n = \frac{n_2 - n_1}{\Delta t_n}.
\end{equation}

One then calculates the Shannon information in order to quantify, how much the light curve is smeared out:
\begin{equation}
  S = \sum_n \frac{w_n}{W} \log\frac{w_n}{W},
\end{equation}
where
\begin{equation}
  W = \sum_n w_n.
\end{equation}
A more narrowly peaked light curve will lead to a larger Shannon information than a more smeared out one.
By varying~$\vartheta$ one then finds that value of the parameter, $\hat\vartheta$, which maximizes~$S$.
Then, $\hat\vartheta$ is considered the best fit value.

In general, the method above will find a value $\hat\vartheta \neq 0$, even if there was no Lorentz-invariance violation at all, due to statistical fluctuations.
In order to determine the significance of this deviation from the null-hypothesis and to determine upper and lower limits on~$\vartheta$, the method described above was repeated on randomized light curves.
For each source~$10^6$ random light curves were produced by keeping all photon arrival times and energies, but assigning to each arrival time a random photon energy out of the set of detected energies (using each energy only once).
We then applied the \emph{DisCan} method to each of these random light curves.
The lower and upper limits on~$\vartheta$ were then determined as single-sided \SI{95}{\percent} confidence limits.

\section{Source selection and data set}\label{sec:sources}
Constraining all 25 non-birefringent LIV parameters of mass-dimension $d=6$ in the SME requires observation of at least 25 astrophysical sources.
To date, limits have been published from 4 VHE AGN~\cite{f_aharonian_etal_2008,*a_abramowski_etal_2011,a_abramowski_etal_2015,j_albert_etal_2008,sd_biller_etal_1999}, 4 GRBs detected with Fermi-LAT~\cite{v_vasileiou_etal_2013}, and one GRB observed by RHESSI~\cite{se_boggs_etal_2004}.
In this paper we supplement that data set with limits on photon dispersion from 25 blazars observed with the Fermi LAT.

\begin{figure}
  \centering
  \includegraphics[width=\columnwidth]{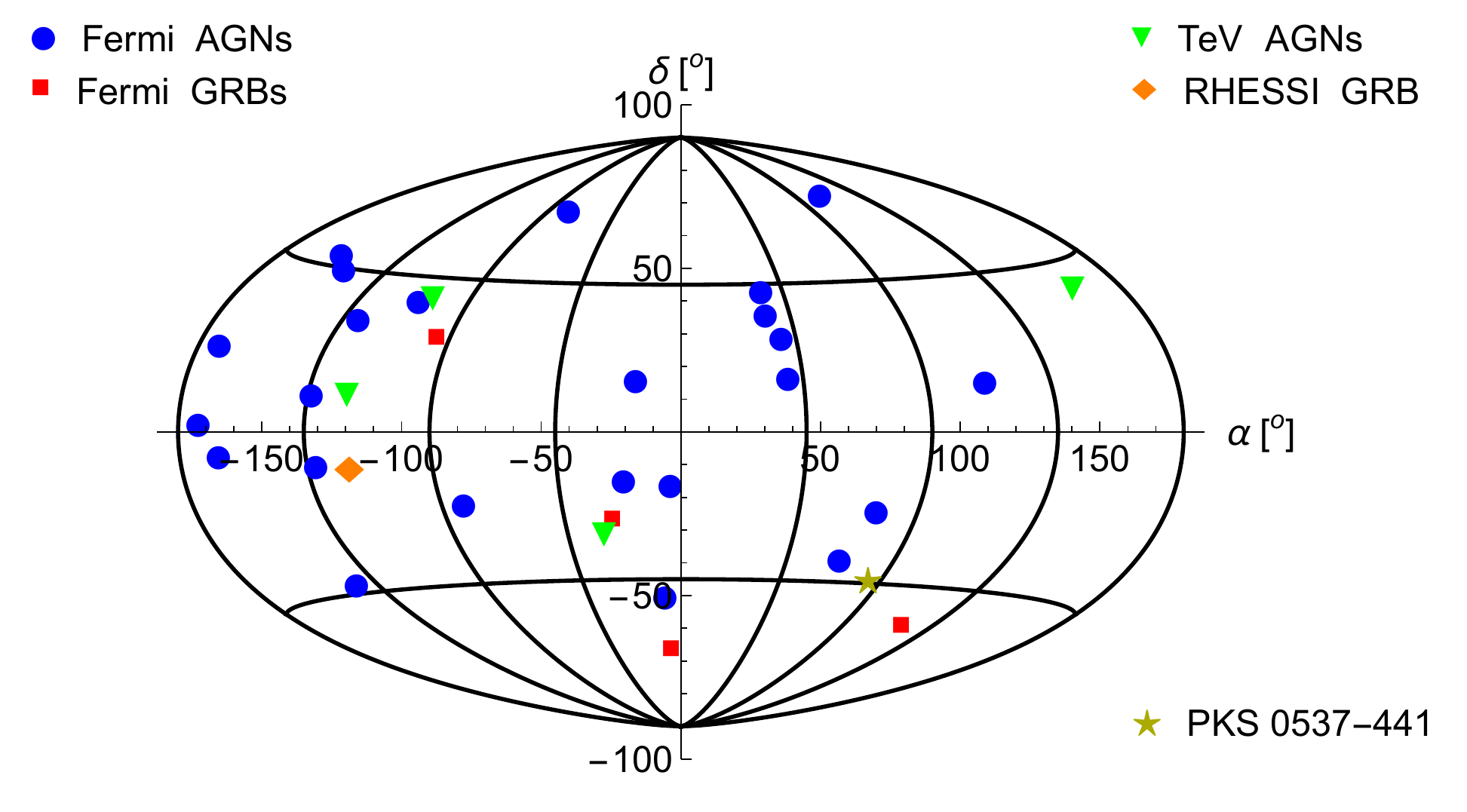}
  \caption{Sky map with all sources used in this analysis in equatorial coordinates. The Fermi AGN (blue circles) and PKS 0527--441 were analyzed in this work.}
  \label{fig:sourcemap}
\end{figure}

We selected the 24 sources from the 4-year Fermi point source catalog (3FGL,~\cite{f_acero_etal_2015}) with the highest variability index~\cite{*[{See }] [{ for the definition.}] pl_nolan_etal_2012}, that also fulfilled the following conditions:
\begin{itemize}
 \item the red shift is known and ${>}0.1$;
 \item no constraints on Lorentz invariance violation have been published based on TeV gamma-ray observations of this source;
 \item the source is significantly detected above \SI{10}{GeV} by Fermi LAT, with $\mathtt{sqrt\_ts\_10\_100\_gev} > 10$ according to the catalog~\cite{pl_nolan_etal_2012}, which is the square root of the logarithm of the likelihood ratio of observing the signal in the $10$ to \SI{100}{GeV} band with and without the point source;
 \item and there are no other similarly bright $\gamma$-ray sources within a $2^\circ$ radius.
\end{itemize}
While the variability index characterizes the variability of a source on month time scales, it does not give any indication about its shorter-term variability, which is of importance to this study.
However, in case of blazars, it is a good indication of the frequency and intensity of observed flares, making it a suitable criterion to select sources for this analysis.
In addition to the sources obtained in this way, we also analyzed the flat spectrum radio quasar S3 0218+35, resulting in a total of 25 blazars, which are listed in Table~\ref{tab:sources}.
Of these, 19 are flat spectrum radio quasars (FSRQs), and 6 are BL Lac type objects.
Figure~\ref{fig:sourcemap} shows a sky map of all sources used in this analysis, including those with previously published limits.

For the analysis of the selected AGNs we used all data from the Fermi Pass 7 \texttt{P7REP\_SOURCE\_V15} data set taken between August 5th, 2008, and October 15th, 2014.
For each source we selected time intervals such that a search radius of \SI{15}{\degree} would pass a zenith angle cut of $\mathtt{zmax} = 100$.
Furthermore, we restricted the energy range to \SI{500}{MeV}--\SI{300}{GeV} and applied the cut \texttt{(DATA\_QUAL>0) \&\& (LAT\_CONFIG==1)}.
Below~\SI{500}{MeV} the direction resolution degrades rapidly and the background contribution rises quickly.
We then applied the \emph{DisCan} method as described in section~\ref{sec:discan} to all remaining events within a search radius of \SI{1}{\degree} around the source.
The results of which will be discussed in the next section.

\begin{table*}[p]
  \centering
  \caption{List of sources studied in this analysis. All source coordinates were obtained from the SIMBAD database~\cite{m_wenger_etal_2000}. Individual references are given for the red shifts.}
  \renewcommand{\tabcolsep}{.5em}
  \begin{tabular}{
      l
      l
      S[table-format=3.3]
      S[table-format=+3.3,retain-explicit-plus]
      S[table-format=2.3]
      l
    }
    \toprule
      \multicolumn{1}{c}{\textbf{Source}}            &
      \multicolumn{1}{c}{\textbf{Class}}             &
      \multicolumn{1}{c}{\textbf{RA}}                &
      \multicolumn{1}{c}{\textbf{Declination}}       &
      \multicolumn{1}{c}{\textbf{Red shift}}         &
      \multicolumn{1}{c}{\textbf{Refs.}}             \\
                                                     &
                                                     &
      \multicolumn{1}{c}{J2000 [${}^\circ$]}         &
      \multicolumn{1}{c}{J2000 [${}^\circ$]}         &
      \multicolumn{1}{c}{$z$}                        &
                                                     \\
    \midrule
      3C 66A        & BL Lac & 35.665  & +43.036 & 0.444 & \cite{d_donato_etal_2001} \\
      3C 273        & FSRQ   & 187.278 &  +2.052 & 0.158 & \cite{k_enya_etal_2002} \\
      3C 279        & FSRQ   & 194.047 &  -5.789 & 0.536 & \cite{wa_barkhouse_pb_hall_2001} \\
      3C 454.3      & FSRQ   & 343.491 & +16.142 & 0.859 & \cite{wa_barkhouse_pb_hall_2001} \\
      4C +14.23     & FSRQ   & 111.320 & +14.420 & 1.814 & \cite{se_healey_etal_2008} \\
      4C +28.07     & FSRQ   &  39.468 & +28.802 & 1.207 & \cite{wa_barkhouse_pb_hall_2001} \\
      B2 1520+31    & FSRQ   & 230.542 & +31.737 & 1.487 & \cite{d_sowardsemmerd_etal_2005} \\
      B3 1343+451   & FSRQ   & 206.388 & +44.883 & 2.534 & \cite{jl_richards_etal_2014} \\
      GB 1310+487   & FSRQ   & 198.181 & +48.475 & 0.501 & \cite{se_healey_etal_2008} \\
      PKS 0235+164  & BL Lac &  39.662 & +16.616 & 0.94  & \cite{iag_snellen_etal_2002} \\
      PKS 0426--380 & BL Lac &  67.168 & -37.939 & 1.030 & \cite{d_donato_etal_2001} \\
      PKS 0454--234 & FSRQ   &  74.263 & -23.414 & 1.003 & \cite{wa_barkhouse_pb_hall_2001} \\
      PKS 0537--441 & BL Lac &  84.710 & -44.086 & 0.896 & \cite{d_donato_etal_2001} \\
      PKS 0716+714  & BL Lac & 110.473 & +71.343 & 0.300 & \cite{d_donato_etal_2001} \\
      PKS 1222+216  & FSRQ   & 186.227 & +21.380 & 0.435 & \cite{wa_barkhouse_pb_hall_2001} \\
      PKS 1424--41  & FSRQ   & 216.985 & -42.105 & 1.522 & \cite{eb_fomalont_etal_2000} \\
      PKS 1502+106  & FSRQ   & 226.104 & +10.494 & 1.838 & \cite{pc_hewitt_v_wild_2010} \\
      PKS 1510--089 & FSRQ   & 228.211 &  -9.100 & 0.361 & \cite{d_donato_etal_2001} \\
      PKS 1633+382  & FSRQ   & 248.815 & +38.135 & 1.814 & \cite{jk_adelmanmccarthy_etal_2008} \\
      PKS 1830--211 & FSRQ   & 278.416 & -21.061 & 2.507 & \cite{jej_lovell_etal_1998} \\
      PKS 2233--148 & BL Lac & 339.142 & -14.556 & 0.609 & \cite{yy_kovalev_etal_1999} \\
      PKS 2326--502 & FSRQ   & 352.337 & -49.928 & 0.518 & \cite{a_hewitt_g_burbidge_1989} \\
      PMN J2345--1555 & FSRQ & 356.302 & -15.919 & 0.621 & \cite{se_healey_etal_2008} \\
      S3 0218+35    & FSRQ   &  35.273 & +35.937 & 0.685 & \cite{n_kanekar_jn_chengalur_2002} \\
      S4 1849+67    & FSRQ   & 282.317 & +67.095 & 0.657 & \cite{jd_linford_etal_2012} \\
    \bottomrule
  \end{tabular}
  \label{tab:sources}
\end{table*}

\begin{table*}[p]
  \caption{Published limits on the redshift and light-travel-time weighted dispersion coefficient $\gamma_k$. Note that in case of the MAGIC result (indicated by a $^*$) a marginal detection was quoted. We converted this result into \SI{95}{\percent} limits as discussed in Section~\ref{sec:limits}.}
  \begin{tabular}{
      l
      l
      S[table-format=3.3]
      S[table-format=+2.3,retain-explicit-plus]
      S[table-format=1.3]
      S[table-format=+3.2e+4]
      S[table-format=3.2e+4]
      l
    }
    \toprule
      \multicolumn{1}{c}{\textbf{Source}}                  &
      \multicolumn{1}{c}{\textbf{Instrument}}              &
      \multicolumn{1}{c}{\textbf{RA}}                      &
      \multicolumn{1}{c}{\textbf{Declination}}             &
      \multicolumn{1}{c}{\textbf{Red shift}}               &
      \multicolumn{1}{c}{$\bm{\gamma}_\mathbf{min}$} &
      \multicolumn{1}{c}{$\bm{\gamma}_\mathbf{max}$} &
      \multicolumn{1}{c}{\textbf{Refs.}}                   \\
                                                           & 
                                                           &
      \multicolumn{1}{c}{J2000 [${}^\circ$]}               &
      \multicolumn{1}{c}{J2000 [${}^\circ$]}               &
      \multicolumn{1}{c}{$z$}                              &
      \multicolumn{1}{c}{[\si{GeV^{-2}}]}                  &
      \multicolumn{1}{c}{[\si{GeV^{-2}}]}                  &
                                                           \\
    \midrule
      GRB 080916C  & Fermi LAT & 119.847 & -56.638 & 4.35  & -8.7e-20 & 2.0e-19 &
        \cite{v_vasileiou_etal_2013} \\
      GRB 090510   & Fermi LAT & 333.553 & -26.597 & 0.903 & -3.1e-21 & 1.6e-21 &
        \cite{v_vasileiou_etal_2013} \\
      GRB 090902B  & Fermi LAT & 264.939 & +27.324 & 1.822 & -3.4e-20 & 5.2e-20 &
        \cite{v_vasileiou_etal_2013} \\
      GRB 090926A  & Fermi LAT & 353.401 & -66.323 & 2.107 & -1.1e-19 & 5.2e-20 &
        \cite{v_vasileiou_etal_2013} \\
      GRB 021206   & RHESSI    & 240.195 &  -9.710 & 0.3   & -1.0e-16 & 1.0e-16 &
        \cite{se_boggs_etal_2004,va_kostelecky_n_russell_2014} \\
      PKS 2155-304 & H.E.S.S.  & 329.717 & -30.226 & 0.116 & -7.4e-22 & 7.4e-22 &
        \cite{f_aharonian_etal_2008,*a_abramowski_etal_2011} \\
      PG 1553+113  & H.E.S.S.  & 238.929 & +11.190 & 0.49  & -5.37e-21 & 3.46e-21 &
        \cite{a_abramowski_etal_2015} \\
      Mrk 501      & MAGIC     & 253.468 & +39.760 & 0.034 & -5.8e-22 & 5.8e-22 &
        \cite{j_albert_etal_2008}$^*$ \\
      Mrk 421      & Whipple   & 166.114 & +38.209 & 0.031 & -1.4e-21 & 1.4e-21 &
        \cite{sd_biller_etal_1999} \\
    \bottomrule
  \end{tabular}
  \label{tab:published-limits}
\end{table*}

\setlength{\rotFPtop}{0pt plus 1fil}
\setlength{\rotFPbot}{0pt}
\begin{sidewaystable*}[p]
  \centering
  \caption{Best fit values of the dispersion coefficient~$\vartheta$, as well as upper and lower limits on $\vartheta$ and redshift and light-travel-time weighted dispersion coefficient~$\gamma$ for the sources studied in this analysis.}
  \renewcommand{\tabcolsep}{.5em}
  \begin{tabular}{
      l
      S[table-format=+6.2]
      S[table-format=+7.2]
      S[table-format=7.2]
      S[table-format=+3.3e+4]
      S[table-format=3.3e+4]
    }
    \toprule
      \multicolumn{1}{c}{\textbf{Source}}                     &
      \multicolumn{1}{c}{$\bm{\hat\vartheta}$}          &
      \multicolumn{1}{c}{$\bm{\vartheta}_\mathbf{min}$} &
      \multicolumn{1}{c}{$\bm{\vartheta}_\mathbf{max}$} &
      \multicolumn{1}{c}{$\bm{\gamma}_\mathbf{min}$}    &
      \multicolumn{1}{c}{$\bm{\gamma}_\mathbf{max}$}    \\
                                                              &
      \multicolumn{1}{c}{[\si{s/GeV^2}]}                      &
      \multicolumn{1}{c}{[\si{s/GeV^2}]}                      &
      \multicolumn{1}{c}{[\si{s/GeV^2}]}                      &
      \multicolumn{1}{c}{[\si{GeV^{-2}}]}                     &
      \multicolumn{1}{c}{[\si{GeV^{-2}}]}                     \\
    \midrule
      3C 66A          &   428    &  -7148    & 11017    & -2.77e-14 & 4.26e-14 \\
      3C 273          &     5.0  &    -19.3  &    24.9  & -2.50e-16 & 3.23e-16 \\
      3C 279          &     2.1  &     -3.8  &     6.3  & -1.16e-17 & 1.92e-17 \\
      3C 454.3        &     0.22 &     -0.20 &     0.65 & -3.31e-19 & 1.08e-18 \\
      4C +14.23       &   843    & -12243    & 12107    & -7.37e-15 & 7.29e-15 \\
      4C +28.07       & -1554    & -22276    & 19750    & -2.34e-14 & 2.07e-14 \\
      B2 1520+31      &    -0.78 &  -1064    &   306    & -8.41e-16 & 2.42e-16 \\
      B3 1343+451     &  -733    & -10753    & 10258    & -4.10e-15 & 3.91e-15 \\
      GB 1310+487     &    55.0  &   -142    &   838    & -4.72e-16 & 2.79e-15 \\
      PKS 0235+164    &   473    &  -3974    & 10628    & -5.84e-15 & 1.56e-14 \\
      PKS 0426--380   &    -0.13 &   -779    &   922    & -1.01e-15 & 1.20e-15 \\
      PKS 0454--234   &  -472    &  -1131    &  2975    & -1.52e-15 & 4.01e-15 \\
      PKS 0537--441${}^\dag$ & 36686    &  -2692    &  2504    &    &          \\
      PKS 0716+714    &  -266    &   -378    &    30.9  & -2.35e-15 & 1.92e-16 \\
      PKS 1222+216    &     0.14 &     -1.9  &     3.9  & -7.54e-18 & 1.55e-17 \\
      PKS 1424--41    &    -7.9  &    -52.1  &    45.5  & -3.99e-17 & 3.48e-17 \\
      PKS 1502+106    &    -1.9  &    -19.4  &    51.2  & -1.15e-17 & 3.04e-17 \\
      PKS 1510--089   &     0.65 &     -3.4  &     5.4  & -1.69e-17 & 2.69e-17 \\
      PKS 1633+382    &     4.0  &  -4891    &  4400    & -2.95e-15 & 2.65e-15 \\
      PKS 1830--211   &   -82.4  &   -129    &   106    & -4.99e-17 & 4.10e-17 \\
      PKS 2233--148   &     0.79 &  -3281    &  2104    & -8.52e-15 & 5.46e-15 \\
      PKS 2326--502   &    -1.0  &   -203    &  1507    & -6.48e-16 & 4.81e-15 \\
      PMN J2345--1555 &    -0.33 &  -2655    &  3390    & -6.72e-15 & 8.58e-15 \\
      S3 0218+35      &    -4.0  &   -125    &    61.1  & -6.33e-15 & 5.96e-15 \\
      S4 1849+67      &    90.4  &  -2837    &  2671    & -2.94e-16 & 1.44e-16 \\
    \bottomrule
  \end{tabular}\\[1ex]
  \begin{minipage}{13.5cm}
    \flushleft
    ${}^\dag$ The value of $\hat\vartheta$ obtained for PKS 0537--441 is far outside the limits set by $\vartheta_\text{min}$ and $\vartheta_\text{max}$.
    As discussed in Section~\ref{sec:limits}, we discarded this source for the remainder of the analysis.
  \end{minipage}
  \label{tab:our-limits}
\end{sidewaystable*}

\section{Limits on LIV parameters}\label{sec:limits}
\subsection{Previously published constraints}
\begin{figure}
  \centering
  \includegraphics[width=.8\columnwidth]{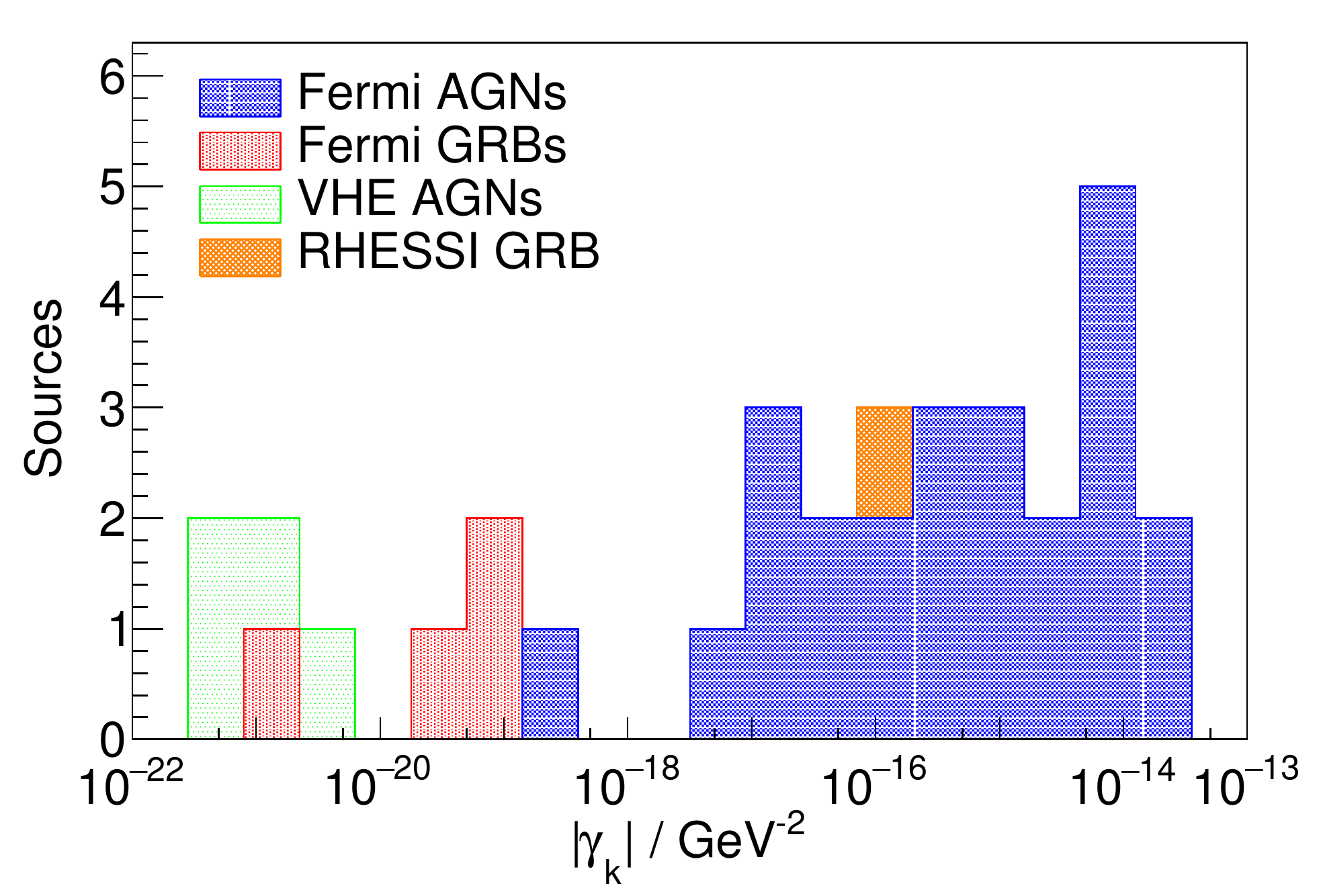}
  \caption{Distribution of weighted dispersion coefficients~$\gamma$ values used in this study, showing the best 95\% confidence level limit obtained for each source. All Fermi AGN were analyzed here, while the values from other sources have previously been published~\cite{v_vasileiou_etal_2013,se_boggs_etal_2004,f_aharonian_etal_2008,j_albert_etal_2008,sd_biller_etal_1999,a_abramowski_etal_2011,a_abramowski_etal_2015}. The Fermi AGN limits are not as sensitive as the Fermi GRB or VHE AGN limits, but they provide the data points required for constraining all 25 expansion coefficients~$c_{(I)jm}^{(6)}$.}
  \label{fig:sources_gamma}
\end{figure}

Previously published results on quadratic photon dispersion are listed in Table~\ref{tab:published-limits}, and the magnitudes of all values including ours are shown in Fig.~\ref{fig:sources_gamma}.
The best limits are obtained from AGNs observed with VHE gamma-ray instruments.
The extremely high photon energies more than compensate for the fact that the observed objects have a relatively low redshift, in particular given that the expected photon dispersion is proportional to the square of the photon energy.
Note, however, that MAGIC originally reported a marginal detection~\cite{j_albert_etal_2008} of $\hat\vartheta = \SI{3.71 \pm 2.57 e-6}{s/GeV^2}$.
Due to the low significance of the result, we decided to convert it into a \SI{95}{\percent} upper limit and then conservatively used the negative value as a lower limit, i.\,e. $\SI{-8.85e-6}{s/GeV^2} \leq \vartheta \leq \SI{8.85e-6}{s/GeV^2}$.
The resulting limits on $\gamma$ are given in the table.
In case of Fermi GRB observations the extremely short temporal structure and high redshifts lead to limits on the order of $\SI[retain-unity-mantissa=false]{1e-19}{GeV^{-2}}$ or better.
The RHESSI GRB limit suffers from the lower attainable energies.

\begin{figure}
  \centering
  \includegraphics[width=\columnwidth]{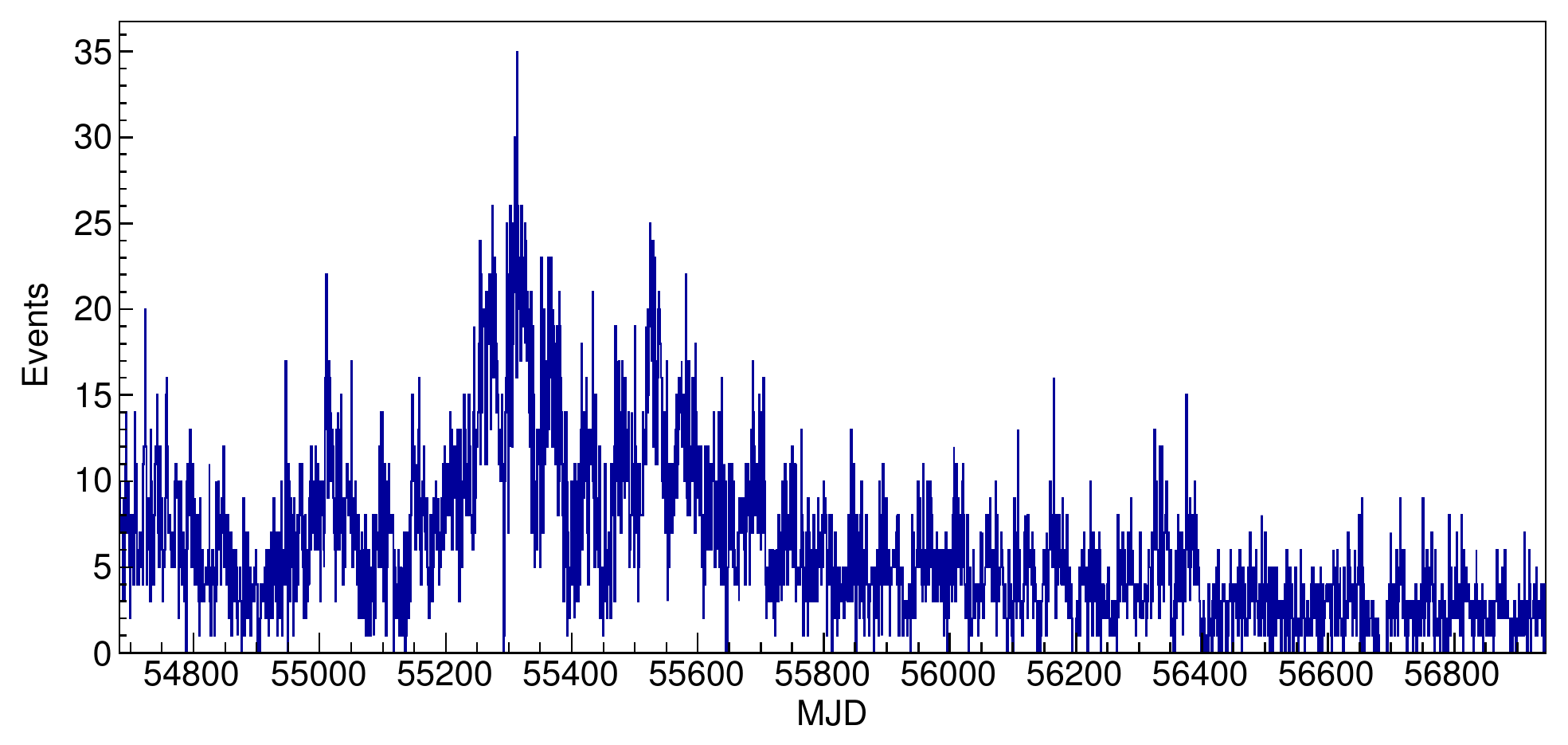}
  \caption{Light curve of PKS 0537--441 above \SI{1}{GeV} as observed with the Fermi LAT. The raw event counts were not exposure corrected reflecting the way individual photons are used in the \emph{DisCan} method.}
  \label{fig:pks0537-441_lightcurve}
\end{figure}

\subsection{Constraints from AGN observed with Fermi}
In this work, we obtained limits on quadratic photon dispersion from an analysis of Fermi AGNs.
As seen in Table~\ref{tab:our-limits}, most of the limits we obtained are between~$10^{-14}$ and $10^{-18}\,\mathrm{GeV^{-2}}$, mostly depending on the brightness of the source and the duration of the observed bursts.
A bright short flare will be sensitive to time structures that can be orders of magnitude smaller than what can be tested with a fainter and longer flare.
This is the main advantage of GRB observations over AGNs.
No significant photon dispersion was found in any of the AGN studied here.
The value of $\hat\vartheta = \SI{36686}{s/GeV^2}$ found in case of the FSRQ PKS 0537--441 is significantly beyond the 95\% upper limit obtained from randomized light curves.
Using these randomized light curves, we determined that the chance probability of observing this value of $\hat\vartheta$ or larger, is less than~0.5\%.
While this by itself would be a significant deviation from the null hypothesis, considering that we analyzed 25 sources, the post-trial probability of this event is~19.2\%.
At the same time, this value of~$\hat\vartheta$ is also significantly larger than any other previously published limit, as well as all constraints found in this analysis, which strongly suggests that this finding is due to a source-intrinsic effect.
Therefore, we removed PKS 0537--441 from our data set, and completed the analysis with the remaining sources.
A more detailed discussion of this source follows in Section~\ref{sub:pks0537-441}.

\subsection{Constraining SME parameters}
The redshift and light-travel-time weighted dispersion coefficients~$\gamma$ are related to the Lorentz violating coefficients~$c_{(I)jm}^{(6)}$ through Eq.~\eqref{eq:cI_system}.
This system of equations can be written in matrix form,
\begin{equation}
  \mathbf{H} \bullet \boldsymbol{v} = \boldsymbol{\gamma},
\end{equation}
where $\boldsymbol{v}$ is a vector of the 25 independent real numbers entering the complex coefficients~$c_{(I)jm}^{(6)}$, $\boldsymbol{\gamma}$ is a vector of the $N$ values of $\gamma_k$ for $N$ sources studied, and $\mathbf{H}$ is the $25 \times N$ matrix relating the two sets, whose rows are obtained directly from Eq.~\eqref{eq:cI_system}.
For a set of uncorrelated measurements of $\gamma_k$ from $N \geq 25$ sources, the best fit set of parameters~$\boldsymbol{v}$ can be obtained through,
\begin{equation}\label{eq:cI_system-inverse}
  \boldsymbol{v} = (\mathbf{H}^T\mathbf{H})^{-1}\mathbf{H}^T\boldsymbol{\gamma} = \mathbf{H}^+\boldsymbol{\gamma},
\end{equation}
where~$\mathbf{H}^+$ is the Moore-Penrose pseudoinverse of~$\mathbf{H}$~(see e.\,g. Ref.~\cite{a_benisrael_tne_greville_2003}, and note that $\mathbf{H}^+ = \mathbf{H}^{-1}$ for square matrices).

We used Eq.~\eqref{eq:cI_system-inverse} to obtain the 95\% confidence level limits on the LIV coefficients~$c_{(I)jm}^{(6)}$ from the limits on~$\gamma_k$, by generating~$10^7$ random vectors~$\boldsymbol{\gamma}$.
For each source, the probability distribution of~$\gamma_k$ was approximated by an asymmetrical normal distribution with mean $0$ and standard deviations chosen to match the one-sided \SI{95}{\percent} limits in Tables~\ref{tab:published-limits} and~\ref{tab:our-limits}.
For each of these random~$\boldsymbol{\gamma}$ we then solved Eq.~\eqref{eq:cI_system-inverse} and in that way found the distribution for each coefficient in~$\boldsymbol{v}$.
From those resulting distributions we then determined the single sided \SI{95}{\percent} upper and lower bounds.
The resulting limits on all 25 non-birefringent Lorentz violating parameters of mass-dimension $d=6$ of the SME are listed in Table~\ref{tab:liv-parameter-limits}.

\begin{table}
  \centering
  \caption{Limits in units of \si{GeV^{-2}} on all independent LIV parameters $c_{(I)jm}^{(6)}$ obtained in this analysis. The dependent parameters $c_{(I)j-m}^{(6)}$ can be calculated according to Eq.~\eqref{eq:reality_condition}.}
  \setlength\extrarowheight{1ex}
  \begin{tabular}{
    S[table-format=+3.3e+4]
    @{ $<$ }c@{ $<$ }
    S[table-format=2.3e+4]
    }
  \toprule
    -2.705e-14 & $c_{(I)00}^{(6)}$                       & 3.925e-14 \\
    -3.753e-14 & $c_{(I)10}^{(6)}$                       & 2.889e-14 \\
    -2.816e-14 & $\Real\bigl(c_{(I)11}^{(6)}\bigr)$      & 3.574e-14 \\
    -3.299e-15 & $\Imaginary\bigl(c_{(I)11}^{(6)}\bigr)$ & 5.984e-15 \\
    -4.232e-14 & $c_{(I)20}^{(6)}$                       & 3.032e-14 \\
    -1.590e-14 & $\Real\bigl(c_{(I)21}^{(6)}\bigr)$      & 1.043e-14 \\
    -4.412e-14 & $\Imaginary\bigl(c_{(I)21}^{(6)}\bigr)$ & 3.288e-14 \\
    -2.353e-14 & $\Real\bigl(c_{(I)22}^{(6)}\bigr)$      & 3.113e-14 \\
    -5.144e-14 & $\Imaginary\bigl(c_{(I)22}^{(6)}\bigr)$ & 6.634e-14 \\
    -4.823e-14 & $c_{(I)30}^{(6)}$                       & 6.435e-14 \\
    -2.439e-14 & $\Real\bigl(c_{(I)31}^{(6)}\bigr)$      & 1.798e-14 \\
    -2.822e-14 & $\Imaginary\bigl(c_{(I)31}^{(6)}\bigr)$ & 2.078e-14 \\
    -3.125e-14 & $\Real\bigl(c_{(I)32}^{(6)}\bigr)$      & 3.855e-14 \\
    -2.171e-14 & $\Imaginary\bigl(c_{(I)32}^{(6)}\bigr)$ & 1.624e-14 \\
    -3.693e-14 & $\Real\bigl(c_{(I)33}^{(6)}\bigr)$      & 2.943e-14 \\
    -4.216e-14 & $\Imaginary\bigl(c_{(I)33}^{(6)}\bigr)$ & 5.656e-14 \\
    -2.313e-14 & $c_{(I)40}^{(6)}$                       & 2.739e-14 \\
    -9.021e-15 & $\Real\bigl(c_{(I)41}^{(6)}\bigr)$      & 1.131e-14 \\
    -2.953e-14 & $\Imaginary\bigl(c_{(I)41}^{(6)}\bigr)$ & 3.904e-14 \\
    -4.650e-15 & $\Real\bigl(c_{(I)42}^{(6)}\bigr)$      & 6.846e-15 \\
    -2.489e-14 & $\Imaginary\bigl(c_{(I)42}^{(6)}\bigr)$ & 1.961e-14 \\
    -7.276e-15 & $\Real\bigl(c_{(I)43}^{(6)}\bigr)$      & 1.014e-14 \\
    -1.246e-14 & $\Imaginary\bigl(c_{(I)43}^{(6)}\bigr)$ & 1.343e-14 \\
    -3.919e-14 & $\Real\bigl(c_{(I)44}^{(6)}\bigr)$      & 2.923e-14 \\
    -1.801e-14 & $\Imaginary\bigl(c_{(I)44}^{(6)}\bigr)$ & 1.427e-14 \\
  \bottomrule
  \end{tabular}
  \label{tab:liv-parameter-limits}
\end{table}

Previous measurements only considered the isotropic case since not enough sources were available.
The previously published results made use of Fermi-LAT GRB observations and TeV gamma-ray observations of AGN.
The resulting isotropic limits are up to 6 orders of magnitude better than the anisotropic limits presented here~\cite{aa_abdo_etal_2009,v_vasileiou_etal_2013,va_kostelecky_n_russell_2014,f_aharonian_etal_2008,a_abramowski_etal_2011,a_abramowski_etal_2015,j_albert_etal_2008,sd_biller_etal_1999}.
The reason for this big difference is that the results of Eq.~\eqref{eq:cI_system-inverse} are dominated by the worst of the best 25 constraints.
As a consequence, the results of this analysis cannot be improved significantly by simply adding more constraints to the data set.
Major improvements will only be possible when a large number of additional highly constraining observations is made (such as TeV observations of further AGN and GeV observations of gamma-ray bursts).

However, our limits are the first constraints on any complete sector of the SME, and the first direct constraints on any of the parameters~$c_{(I)jm}^{(6)}$ other than~$c_{(I)00}^{(6)}$, which describes the isotropic case~\cite{va_kostelecky_n_russell_2014}.
No Lorentz invariance violation has been observed in the photon dispersion in energy or direction.

\subsection{PKS 0537--441}\label{sub:pks0537-441}
We used the constraints on the coefficients~$c_{(I)jm}^{(6)}$ in order to test to what degree the finding of a non-zero delay of high energy photons from PKS 0537--441 is consistent with quadratic photon dispersion in the SME given the observations of the other sources in this analysis.
In the same way as described above, we generated $10^7$ random vectors~$\boldsymbol{\gamma}$ and then computed the coefficients~$c_{(I)jm}^{(6)}$ according to Eq.~\eqref{eq:cI_system-inverse}.
For each of these sets of parameters we then calculated the expected value of~$\vartheta$ for PKS 0537--441 according to Equations~\eqref{eq:gamma_k} and~\eqref{eq:cI_system}.
In this way we found that based on the constraints on the Lorentz invariance breaking parameters in the SME obtained from the other sources one expects a value of $\hat\vartheta < \SI{15775}{s/GeV^2}$ for PKS 0537--441 at the 95\% confidence level.
Furthermore, the probability of finding $\hat\vartheta \geq \SI{36686}{s/GeV^2}$ is only $7.1 \times 10^{-5}$.
This underlines our earlier conclusion that the result found here has to be caused by a source-intrinsic effect, and not by photon dispersion caused by Lorentz invariance breaking.

The light curve of PKS 0537--441 (Fig.~\ref{fig:pks0537-441_lightcurve}) shows an extended period of high activity between MJD 55253 and 55708 (February 2nd, 2010, and May 27th, 2011), which can be subdivided into at least two major flares.
The first one lasting through MJD 55392 (July 15th, 2010) and the second one starting at MJD 55505 (November 5th, 2010).
We analyzed those two flares independently and found no photon arrival time variation during the first, larger flare.
During the second flare, an arrival time variation comparable to the value found for the entire light curve could be observed.
Assuming that the Lorentz invariance violating coefficients are constant in time, this is a contradiction that suggests that there was a source-intrinsic spectral evolution during the second flare.

\section{Summary}\label{sec:summary}
\begin{figure}
  \centering
  \includegraphics[width=\columnwidth]{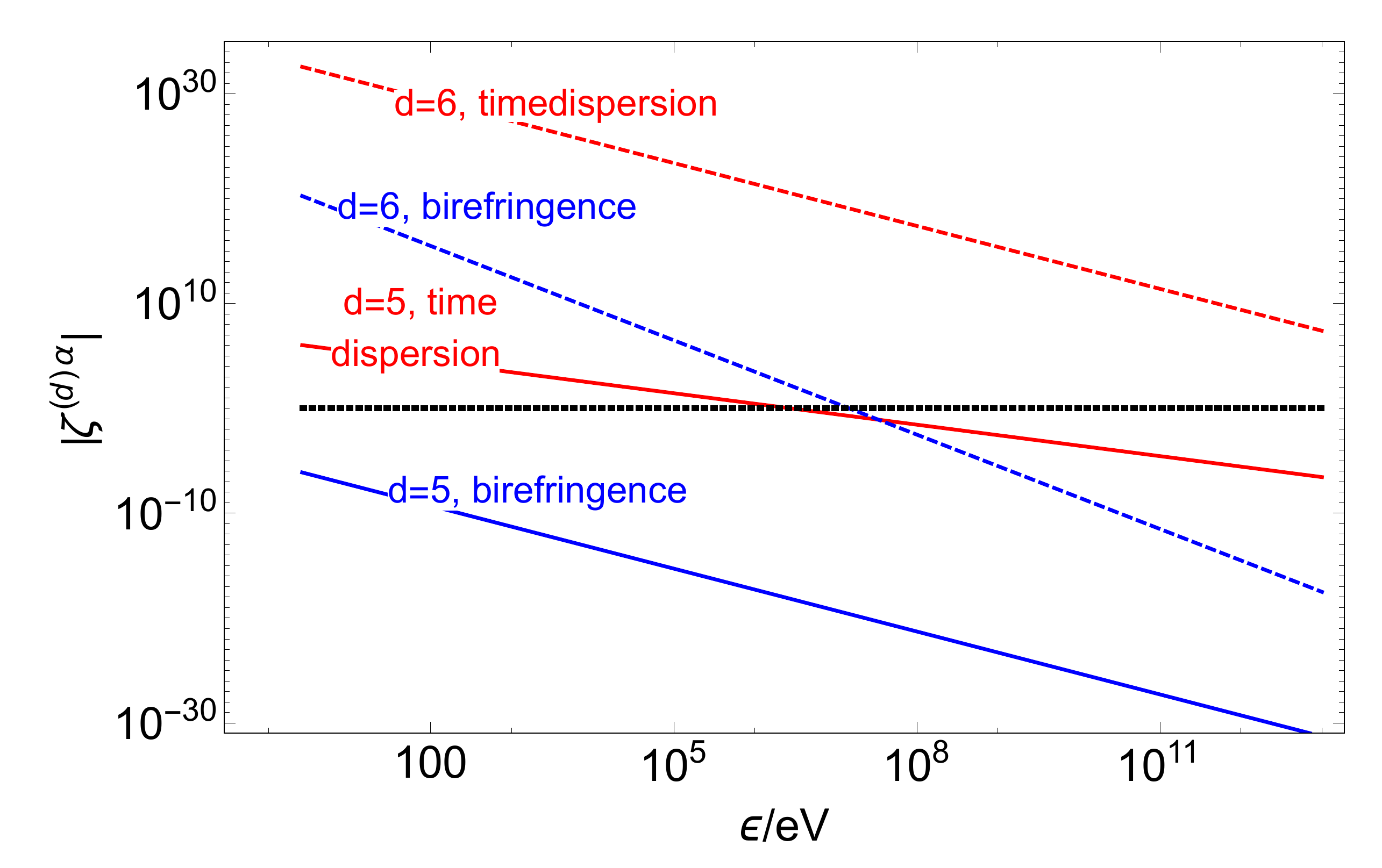}
  \caption{Estimates of the limits on Lorentz invariance violating paramters $\zeta^{(d)a} = E_\text{Planck}^d\varsigma^{(d)a}$ with $d=5$ and $d=6$ and $a \in \{0,3,+\}$ (see Section~\ref{sec:framework}) that can be achieved with future time dispersion and birefringence observations at energy $\epsilon$. Results at and below the dotted line at $\zeta^{(d)a} = 1$ constrain effects at the Planck energy scale.}
  \label{fig:achievable_limits}
\end{figure}

In the Standard-Model Extension, Lorentz invariance violation is described by non-renormalizable terms of mass-dimension $d \geq 5$.
Dimension 5 operators have already been constrained very strongly through polarization measurements.
In addition, these operators not only violate Lorentz symmetry but also CPT making it plausible that CPT-even operators of higher mass-dimension $d=6$ constitute the leading order.
There is a subset of 25 non-birenfringent operators of $d=6$ leading to an anisotropic photon dispersion that is quadratic in energy.
These terms are characterized by a set of 25 real coefficients, which can be constrained through astrophysical dispersion measurements from 25 or more directions in the sky.
We conducted a search for Lorentz violating photon dispersion from 25 Active Galactic Nuclei using data from the Fermi Large Area Telescope (LAT).
Using the \emph{DisCan} method we did not find any significant energy dependence of the speed of light with one exception.
In the case of PKS 0537--441, which exhibited a strong energy dependence of the photon arrival times, we demonstrated that this is most likely a source-intrinsic effect observed during one of its flares and absent at other times.
Therefore, we set upper and lower limits on the coefficients describing the quadratic photon dispersion for all sources.
We combined our 24 limits with 9 previously published constraints in order to set limits on all 25 coefficients of the non-birefringent Lorentz-violating operators of mass-dimension $d=6$ in the Standard-Model Extension.
While previous measurements were able to constrain linear combinations of all operators, our limits represent the first set of constraints on a complete subset of individual coefficients in the photon sector with $d=6$.
The photon sector of the SME has always been the best-constrained part of the theory.
However, the detection of high-energy neutrinos by IceCube promises to provide constraints on the neutrino sector in the near future~\cite{fw_stecker_etal_2015}.

The next step will be to repeat the analysis presented here using polarization data in order to constrain the birefringent coefficients in a similar way.
Polarimetric observations rule out a modification of the photon dispersion relation of order unity at the Planck-scale from operators with $d = 5$ by more than six orders of magnitude.
In contrast, neither time-of-flight measurements nor polarimetric observations do so for the case of $d = 6$.
It is instructive to evaluate how much better future time-lag and polarization measurements will do in this regard.
For this purpose we assume that observations of GRBs at $z = 1$ can constrain the time-of-flight difference of photons of energies $E_1 = \epsilon$ and $E_2 = 0.1\epsilon$ with an accuracy of \SI{1}{ms}, and succeed to detect a polarized signal from these GRBs.
Figure~\ref{fig:achievable_limits} shows the resulting constraints.
Interestingly, the time-of-flight measurements will not have the sensitivity required to constrain new physics at the Planck scale for the case of $d = 6$.
Polarization observation do better, but require the detection of polarized signals at ${>}\SI{20}{MeV}$ energies.
Such detections might be possible with a next-generation Compton or pair production telescope (e.\,g.~\cite{sd_hunter_etal_2014}).

\section*{Acknowledgements}
We would like to thank Alan Kostelecký, Jim Buckley, Floyd Stecker and Manel Errando for fruitful discussions.
The authors are grateful for funding from NASA Grant \#NNX14AD19G and DOE Grant \#DE-FG02-91ER40628.
This research is based on data collected by the Fermi Large Area Telescope made publicly available by the Fermi Science Support Center.
It has made use of the SIMBAD database, operated at CDS, Strasbourg, France.

%%\bibliographystyle{plainnat}
%\nocite{*}  % to test bibliography layout
\bibliography{liv}

\end{document}